\documentclass{emulateapj}

\usepackage{amsmath}

\newcommand{\kms}{km\,s$^{-1}$}

\shorttitle{\sc
A Critical Shock Mach Number for Particle Acceleration: $M=\sqrt{5}$
}
\shortauthors{J. Vink \&  R. Yamazaki}

\begin{document}

\title{\sc
A Critical Shock Mach Number for Particle Acceleration in the Absence of Pre-existing Cosmic Rays: $M=\sqrt{5}$
}
\author{Jacco Vink$^1$ and Ryo Yamazaki$^2$}
\affil{$^1$Astronomical Institute Anton Pannekoek/Gravitation and AstroParticle Physics Amsterdam (GRAPPA), 
University of Amsterdam, Science Park 904, 1098XH Amsterdam, the Netherlands\\
$^2$Department of Physics and Mathematics, College of Science and Engineering, Aoyama Gakuin University, 5-10-1 Fuchinobe, Chuo-ku, Sagamihara, Kanagawa 252-5258, Japan
}

\email{j.vink@uva.nl}

\begin{abstract}
It is shown 
that, under some generic assumptions, shocks cannot accelerate particles
unless the overall shock Mach number exceeds a critical value $M>\sqrt{5}$. The reason is that
for $M\leq \sqrt{5}$ the work done to compress the flow
in a particle precursor requires more enthalpy flux
than the system can sustain. 
This lower limit applies to situations without significant magnetic field pressure. 
In case that the  magnetic field pressure dominates the pressure in the unshocked medium, 
i.e. for low plasma beta, the resistivity of the magnetic field
makes it even more difficult to fulfil the energetic requirements for the 
formation of shock with an accelerated particle precursor and associated
compression of the upstream plasma.
We illustrate the effects of magnetic fields for the extreme situation of
a purely perpendicular 
magnetic field configuration with plasma beta $\beta=0$, which gives a minimum
Mach number of $M=5/2$.
The situation becomes more complex, if we incorporate the effects of pre-existing cosmic rays,
indicating that the additional degree of freedom allows for less strict Mach number limits on acceleration.
We discuss the implications of this result 
for low Mach number shock acceleration as found in solar system shocks, and shocks in clusters of galaxies.
\end{abstract}


\section{Introduction}

Collisionless shock waves
occur in a wide variety of astrophysical settings, and involve a wide
variety of length and energy scales.
Examples are, on the scales of the solar system,
the Earth' bow shock, and the solar wind
termination shock; on parsec scales, supernova remnants shocks;
and on  megaparsec scales, the shocks in clusters of galaxies.

In many cases collisionless shocks are associated with particle acceleration.
It is, for example, generally thought that the origin of Galactic cosmic rays,
with proton  energies up to $3\times 10^{15}$~eV,
are high-Mach-number supernova remnant shocks
\citep{helder12}, 
whereas the ultra-high energy cosmic rays, up to $10^{20}$~eV, are
usually
associated with relativistic shock waves caused by active galactic nuclei,
or gamma-ray bursts \citep{kotera11}.

\begin{figure*}
\centerline{
\includegraphics[width=0.95\textwidth]{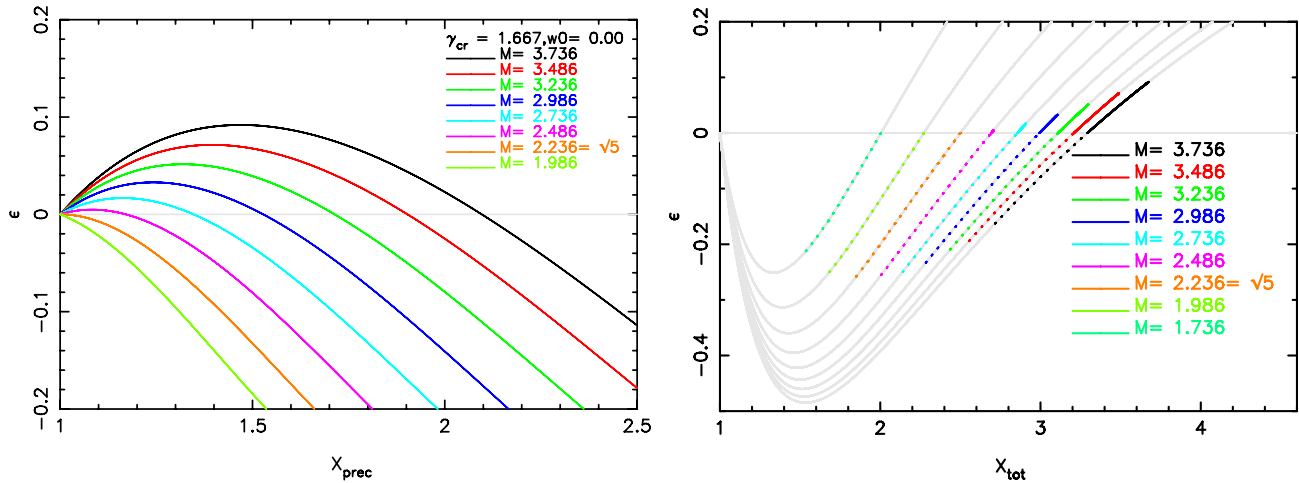}
}
\caption{
Left panel: The curves represent the solutions of the energy-flux escape parameter $\epsilon$ 
as a function of the
precursor compression ratio $\chi_{\rm prec}$, for various values of the 
overall Mach number, with increments of $\Delta M_{\rm g,0}=0.25$.
The slopes at $\chi_{\rm prec}=1$
are negative for $M_{\rm g,0}<\sqrt{5}$, resulting in negative values of $\epsilon$,
which is unphysical. For $M_{\rm g,0}>\sqrt{5}$ one does obtain physical
solutions, but energy escape is required ($\epsilon>0$).
Right panel:
The behavior of $\epsilon$ as a function
of total shock compression $\chi_{\rm tot}$ (Eq.~\ref{eq:epsilon})
for the same Mach numbers as in the left panel.
The total, light grey, 
curve shows
a wide range of shock compression ratios, but only values $\epsilon\ge 0$ correspond
to potientially physical solutions.
The colored curves are solutions to the two-fluid model of 
\citet{vink10a}, with the unphysical solutions ($\chi_{\rm prec}<1$) 
indicated with a dotted line. 
The highest values of $\chi_{\rm tot}$ of the colored lines correspond to the maximum
compression ratios as given by Eq.~\ref{eq:chitot}.
The compression ratios with $\epsilon=0$ correspond 
to the standard Rankine-Hugoniot solutions.
\label{fig:betalarge}
}
\end{figure*}

Low Mach number shocks are also associated with particle acceleration,
but not always.
For example, some shocks driven by coronal mass ejections (CMEs),
which have magnetosonic Mach numbers $M_{\rm ms}\lesssim 4$,
are accompanied by Type II radio burst \citep[e.g.][]{gopalswamy10}, 
whereas others are not. Type II radio bursts are often considered a sign
for particle acceleration. 
The solar wind termination shock has a similarly low Mach number, of around
2.5 \citep{lee09}, and is associated with particle
acceleration \citep[e.g.][]{florinski09}.
On a much larger scale, some shocks in clusters of galaxies result
in so-called radio relics, elongated  structures that emit radio synchrotron
emission \citep[e.g.][]{vanweeren10}. But not
all cluster shocks identified in X-rays 
appear to be accompanied by radio emission.
The typical shock velocities
in clusters of galaxies are of the order of a few 1000~\kms.
But due
to the high temperatures, and hence high sounds speeds, of the
plasma in which the shocks propagate, 
the Mach numbers are modest, with
$M_{\rm ms}\lesssim 3$ \citep{markevitch07}.

In many cases particle acceleration by shocks is attributed to diffusive 
shock acceleration \citep[][for a review]{malkov01}. 
According to the diffusive shock acceleration theory, elastic scattering of energetic,
charged particles 
on both sides of the shock 
causes particles to cross the shock front repeatedly. Each shock crossing results in an average
increase in momentum of order $\Delta p/p\sim V_{\rm s}/c$, 
with $V_{\rm s}$ the shock velocity, and $c$ the speed of light.
The scattering of the particles is caused by magnetic
field fluctuations/plasma waves. The interaction of these particles
with the magnetic field fluctuations causes the accelerated particles
to exert a pressure on the upstream plasma (i.e. the unshocked medium), which
results in the formation of a shock precursor that compresses and slows down
the plasma before it enters the actual shock (which is labeled subshock,
in order to distinguish it from the total shock structure). 
This back-reaction of
the shock-accelerated particles on the plasma flow has been observed in-situ
at the solar termination shock, as measured by Voyager 2 \citep{florinski09}.

The purpose of this paper is to show that particle acceleration, under
general assumptions, requires a minimum Mach number of $M=\sqrt{5}$,
and somewhat higher if magnetic fields are dynamically important
(i.e. for low plasma betas, with $\beta\equiv 8\pi nk_{\rm B}T/B^2<1$).

Note that the  critical Mach number discussed here
is distinct from the so-called first critical Mach number, $M_{\rm c}$,
which is often mentioned in the literature on collisionless shocks 
\citep{marshall55,edmiston84,treumann09}. The first critical Mach number
concerns the details of the shock formation process itself in the presence of magnetic fields.
The magnetic pressure component prevents shocks with Mach numbers lower
than the critical Mach number to heat the post-shock plasma to temperature
where the flow-speed is subsonic. Similar  critical Mach numbers exist for
shocks moving through a medium with pre-existing cosmic rays \citep{becker01}.

The critical Mach number discussed in this paper concerns the overall 
thermodynamic properties of shocks with a precursor of accelerated particles. 
In order to
explain it, we draw upon the two-fluid model of \citet{vink10a}.
In this paper it was already noted that particle acceleration seemed impossible
for low Mach numbers, but the exact Mach number was not given.
In addition, we derive here the critical Mach number for acceleration 
for perpendicular shocks with $\beta=0$, and discuss the more peculiar
case when there are pre-existing cosmic rays.

\section{A minimum Mach number for diffusive shock acceleration}
\label{sec:highbeta}

\subsection{The Rankine-Hugoniot relations extended with a cosmic-ray component}
Shock jump conditions  are  governed by the so-called Rankine-Hugoniot relations
\citep[e.g.][]{zeldovich66,tidman71},  which describe the state of the media on
both sides of the shock, based on the equation of state and the conservation
of mass-, momentum, and energy-flux. These equations assume, therefore, steady state conditions. 

Non-linear particle acceleration \citep{malkov01}, however, may change shock-jump conditions in astrophysical
shocks, as the pressure of particles in the shock precursor compresses the plasma flowing into
the shock, and because the highest energy particles may escape the shock region.
The escape of the highest energy particles
does hardly affect mass- and momentum-flux conservation across the whole shock region, 
since only a very small fraction of the particles escape, but it does violate energy-flux conservation, as the escaping
particles are typically particles that have gained considerable energy \citep{berezhko99}.
Some of the physics of non-linear particle acceleration can be captured by treating the accelerated
particles as a separate component, which is referred to as a two-fluid model \citep[e.g.][]{drury81}.  
The accelerated particles contribute to the  pressure on both sides of the subshock. Since the length scale associated
with the subshock is small compared to gradient over which the accelerated particle pressure changes,
the accelerated particles do not change the properties of the subshock directly,
as the pressures of the accelerated particles just upstream and downstream of the shock are equal.
However, the pressure of the accelerated particles  upstream of the subshock 
results in a compression and slowing down of the plasma
flowing into the subshock.
As a result the Mach number just upstream of the subshock
 is smaller than the overall Mach number as measured far upstream.

\citet{vink10a} showed that one can incorporate an accelerated
particle (cosmic-ray) component in the Rankine-Hugoniot relations
by evaluating the Rankine-Hugoniot relations 
in three distinct regions: 
0) the (undisturbed) far upstream medium, 1) in the shock precursor,
just upstream of the subshock, and 2) downstream of the subshock.
The solutions allow for energy to escape from the system, which in kinetic
models for cosmic-ray acceleration is either a result of having particles
remove once they reach a certain maximum momentum \citep[e.g.][]{blasi05},
or by imposing a maximum length scale to which particles are allowed to
diffuse upstream \citep{reville09}. 

In Appendix~\ref{sec:appendixa} the results of the extended Rankine-Hugoniot relations of \cite{vink10a}
are summarized and extended by allowing also for pre-existing cosmic-rays.
The input parameters of the extended Rankine-Hugoniot relations are the
upstream gas Mach number ($M_{\rm g,0}$) and the fractional pressure upstream in cosmic
rays, $w_0=P_{\rm cr,0}/P_{\rm tot}$  (Eq.~\ref{eq:w}). For the cosmic-ray component one has to assume
an adiabatic index, $4/3\leq \gamma_{\rm cr}\leq 5/3$. The extended Rankine-Hugoniot relations
give the downstream pressure contribution of cosmic rays, $w_2$  (Eq.~\ref{eq:w2}), as a function
of the cosmic-ray precursor compression ratio, $\chi_{\rm prec}$ (Eq.~\ref{eq:chi}).
Note that like more elaborate cosmic-ray acceleration models \citep[e.g.][for an overview]{caprioli10},
and the classical two-fluid models \citep{drury81,becker01}, the extended Rankine-Hugoniot
relations assume a steady state situation.

\subsection{A minimum Mach number for acceleration}

The gas flowing into the subshocks behaves like  a standard, classical shock, but due to compression
in the cosmic-ray precursor, the subshock Mach number, $M_{\rm g,1}$, is lower than the upstream
Mach number $M_{\rm g,0}$. The compression ratio at the subshock is given by Eq.~\ref{eq:chisub} in Appendix~\ref{sec:appendixa}.
Since the basic parameter of the extended Rankine-Hugoniot relation is the precursor 
compression ratio $\chi_{\rm prec}$ the total compression ratio for a cosmic-ray accelerating shock is
\begin{equation}
\chi_{\rm tot}=\chi_{\rm prec}\chi_{\rm sub}=
\frac{
(\gamma_{\rm g} + 1)M_{\rm g,0}^2\chi_{\rm prec}^{-\gamma_{\rm g}}}{
(\gamma_{\rm g}-1)M_{\rm g,0}^2\chi_{\rm prec}^{-(\gamma_{\rm g}+1)} + 2}.\label{eq:chitot}
\end{equation}
According to  Eq.~\ref{eq:chitot} the total compression ratio can be larger than
that allowed by standard shock jump relation\footnote{See Eq.~\ref{eq:chisub}, but in this case changing the subscript "sub" by "tot".}
as long as Eq.~\ref{eq:energy} is obeyed, with $\epsilon > 0$ \citep[see also][]{berezhko99}.

The maximum value for the compression ratio can be found by 
solving $d\chi_{\rm tot}/d\chi_{\rm prec}=0$, with
$\chi_{\rm tot}$ given by Eq.~\ref{eq:chitot}.
This shows that the maximum total compression ratio
occurs for
\begin{equation}
\chi_{\rm prec}=
\left( 
\frac{(\gamma_{\rm g}-1)}{2\gamma_{\rm g}}M_{\rm g,0}^2
\right)^{1/(\gamma_{\rm g}+1)}=\left(\frac{1}{5} M_{\rm g,0}^2\right)^{3/8},
\label{eq:chiprecmax}
\end{equation}
with $\gamma_{\rm g}=5/3$.
By inserting Eq.~\ref{eq:chiprecmax} in Eq.~\ref{eq:chitot}
one finds the corresponding sub-shock compression ratio
\begin{equation}
\chi_{\rm sub}=\frac{\gamma_{\rm g}}{\gamma_{\rm g}-1}=\frac{5}{2},
\label{eq:chisubmax}
\end{equation}
which, according to Eq.~\ref{eq:chisub} corresponds to $M_{\rm g,1}=\sqrt{5}$.

This result was obtained by \citet{vink10a}, but an important aspect 
for shocks without pre-existing cosmic-rays (i.e. $w_0=0$)
was not recognized:
Eq.~\ref{eq:chiprecmax} indicates that the solution becomes unphysical
for $M_{\rm g,0}<\sqrt{5}$ as it requires a rarefaction instead of a compression in the cosmic-ray precursor
($\chi_{\rm prec}<1$). So below  $M_{\rm g,0}<\sqrt{5}$ the only
allowed solution is one in which there is no cosmic-ray precursor, and for which
the compression ratio is given by the standard Rankine-Hugoniot relations.

We refer to this critical Mach number as $M_{\rm acc}$, in order to distinguish it
from the first critical Mach number, $M_{\rm c}$ \citep{edmiston84}, and the related critical Mach
numbers investigated by \citet{becker01}.
As we will describe below, for shocks moving through a magnetized medium (section~\ref{sec:beta}).,
or for a (partially) relativistic cosmic-ray population ($\gamma_{\rm cr}<5/3$, section~\ref{sec:relativistic}) $M_{\rm acc}> \sqrt{5}$.
However, as we will discuss in section~\ref{sec:w0}, a population of pre-existing cosmic rays, may
result in cosmic-ray acceleration for values lower than $M_{\rm acc}$.

The maximum value for the energy flux escape, $\epsilon$,  is determined
by solving 
$d\epsilon/d\chi_{\rm prec}=(d\epsilon/d\chi_{\rm tot})(d\chi_{\rm tot}/d\chi_{\rm prec})=0$.
For $\gamma_{\rm cr}=5/3$ this equation has two possible solutions. One corresponds to
a minimum of $\epsilon$, with $\epsilon <0$.
This minimum does not have a physical meaning.
The other solution corresponds to $d\chi_{\rm tot}/d\chi_{\rm prec}=0$,
and is associated with a maximum value of
$\epsilon$, and hence with the maximum of $\chi_{\rm tot}$ (Eq.~\ref{eq:chiprecmax}).

\begin{figure}
\centerline{
\includegraphics[angle=-90,width=0.5\textwidth]{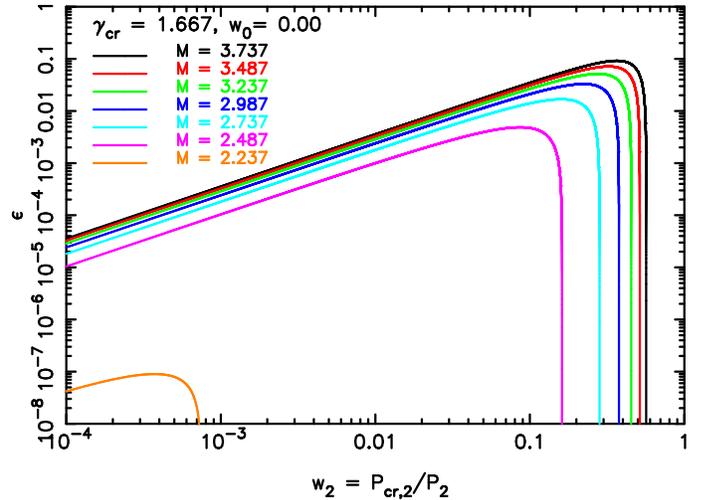}
}
\caption{
The solutions to two-fluid model of \citet{vink10a}. The values for
the Mach number correspond to those in Fig.~\ref{fig:betalarge}, except
that the orange curves correspond to $M_{\rm g,0}=\sqrt{5}+0.001$, in order
to show the behavior very close the critical Mach number.
\label{fig:eps_vs_w}
}
\end{figure}

Fig.~\ref{fig:betalarge} illustrates the properties of
the energy flux equation for shocks with Mach numbers around
$M_{\rm g,0}=\sqrt{5}$ and $\gamma_{\rm cr}=5/3$, indicating that the accelerated particles are non-relativistic. 
The panel on the left shows that
for $M_{\rm g,0}<\sqrt{5}$ and $\chi_{\rm prec}>1$
one obtains $\epsilon<0$, which is unphysical. 
A solution with $\epsilon=0$ is always
possible, and occurs for $\chi_{\rm prec}=1$.
This solution corresponds to  the standard
Rankine-Hugoniot relations.

\begin{figure}
\centerline{
\includegraphics[angle=-90,width=0.5\textwidth]{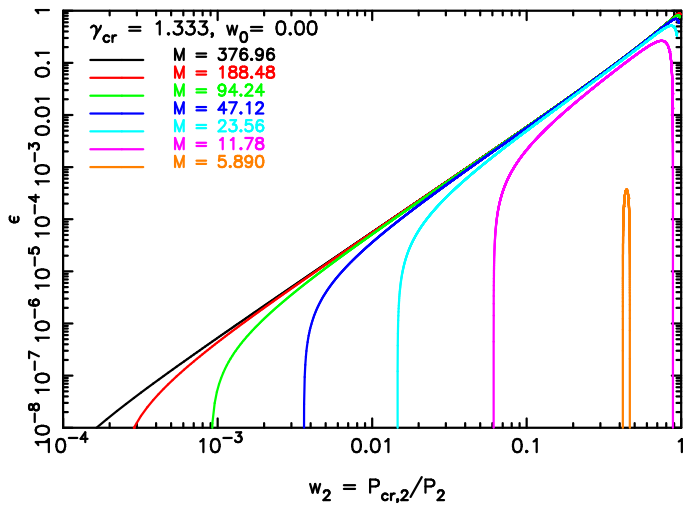}
}
\caption{
The same as Fig.~\ref{fig:eps_vs_w}, but now for an accelerated
particle component characterised by $\gamma_{\rm cr}=4/3$,
for logarithmically spaced intervals of the Mach number.
\label{fig:eps_vs_w_rel}
}
\end{figure}

The right-hand panel of Fig.~\ref{fig:betalarge} shows the behavior of the energy escape ($\epsilon$, Eq.~\ref{eq:epsilon})
 as a function of total compression ratio. Note that this figure does not rely on the details of
 a two-fluid model, as only the total compression ratio is used, but an effective adiabatic index $\gamma$ needs to be specified.
The figure shows that higher compression ratios than the standard shock-jump conditions are allowed, but only if 
there is energy flux escape, i.e. $\epsilon >0$. But in the context of a system with precursor compression
and a subshock, there is a restriction on
the total compression ratios that are possible, namely $\chi_{\rm prec}\geq 1$.
As a consequence, physical solutions
with higher compression ratios  than the standard shock jump conditions
are only possible for $M_{\rm g,0}>\sqrt{5}$. These physical solutions
are indicated by solid colored lines.
 
\begin{figure*}
\centerline{
\includegraphics[width=0.95\textwidth]{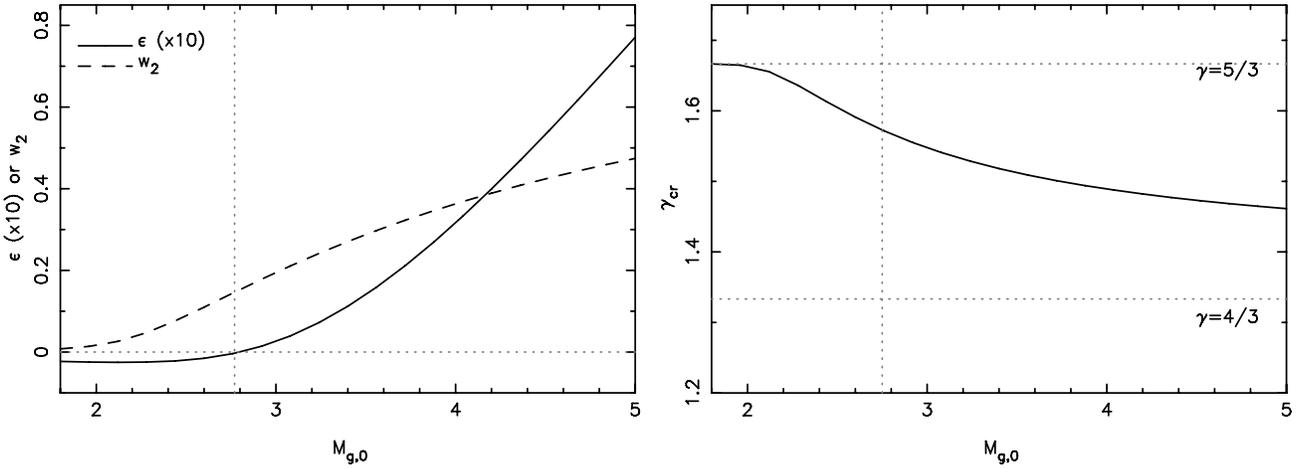}
}
\caption{
The shock solutions as obtained with the kinetic shock acceleration 
model of \citet{blasi05} for a shock velocity of $V_{\rm s}=10$~km\,s$^{-1}$,
and a maximum momentum of the accelerated particles of $p_{\rm max}=100$mc.
Left: the values for the escape flux, $\epsilon$ (multiplied by 10),
and $w_2$. Right: effective adiabatic index $\gamma_{\rm cr}$ of the accelerated
particles. The limiting Mach number for this case is
$M_{\rm acc}\approx 2.79$.
\label{fig:kinetic}
}
\end{figure*}

Fig.~\ref{fig:eps_vs_w} shows the allowed combinations of the fractional downstream cosmic-ray pressure $w_2$ and $\epsilon$.
It illustrates that there is a dramatic change in the maximum possible particle 
acceleration efficiency  going from a Mach number around
$M_{\rm g,0}=2.5$ to a Mach number very close
to $M_{\rm acc}=\sqrt{5}$.

There are other potential effects that may shift the limiting Mach number to higher values.
In section~\ref{sec:beta}, the effects of plasma-beta is treated. But another factor is non-adiabatic heating
in the precursor. Up to now it was assumed that the accelerated particles compress the
upstream plasma, and heats it only adiabatically. 
However, additional heating may occur in the
precursor, for example 
through Coulomb collisions,
wave damping, 
or through friction with neutral atoms \citep{ohira10,raymond11,morlino13}. This leads to higher values of the critical
Mach number. This can be easily seen by replacing Eq.~\ref{eq:m1} by
\begin{equation}
M_{\rm g,1}^2= \frac{M_{\rm g,0}^2\chi_{\rm prec}^{-(\gamma_{\rm g}+1)}}{(1+\alpha)},\label{eq:machsub2}
\end{equation}
with $\alpha\geq 0$ a parameter that parameterizes the additional heating as an
additional fraction of the adiabatic heating, resulting in a lower subshock Mach number.
It can be easily seen that introducing
the additional factor $1/(1+\alpha)$ in Eq.~\ref{eq:chitot} 
results in increasing $M_{\rm acc}$ by a factor $\sqrt{1+\alpha}$.

\subsection{The minimum Mach number for acceleration to a relativistically dominated cosmic-ray population}
\label{sec:relativistic}

In the previous section the limit for particle acceleration was obtained by assuming that the
accelerated particles are non-relativistic ($\gamma_{\rm cr}=5/3$).
This gives the lowest limit on particle acceleration one can obtain. If instead the accelerated particles
are dominated by relativistic particles ($\gamma_{\rm cr}=4/3$), $M_{\rm acc}$ needs to be much higher.
Deriving the value for $M_{\rm acc}$ is much more difficult as the overall equation of state of the
two-fluid plasma depends now on the mixture of thermal particles and accelerated particles.
Instead we give here the numerical value we obtained,  $M_{\rm acc}=5.882$.

\begin{figure}
\centerline{
\includegraphics[angle=-90,width=0.5\textwidth]{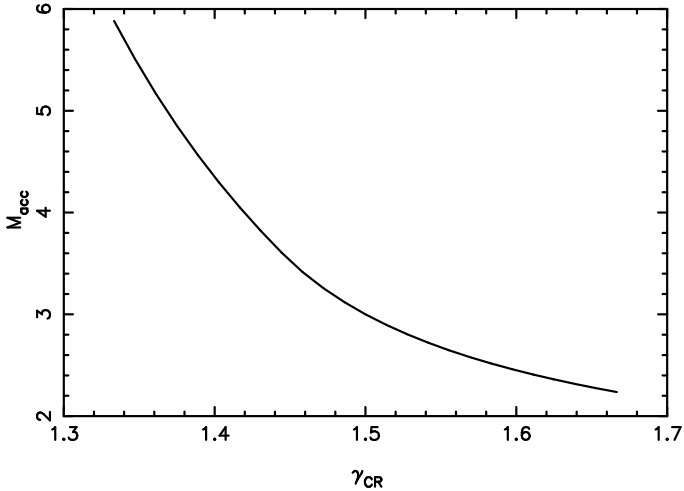}
}
\caption{
The critical Mach number as a function of assumed adiabatic index
for the accelerated particle populuation ($\frac{4}{3}\le \gamma_{\rm cr} \le \frac{5}{3}$).\label{fig:mcrit}
}
\end{figure}

Fig.~\ref{fig:eps_vs_w_rel} shows the behavior of energy  escape and downstream cosmic-ray pressure
for $M_{\rm g,0}> M_{\rm acc}=5.882$. It illustrates a peculiar feature of the solutions for
$\gamma_{\rm cr}=4/3$ as compared $\gamma_{\rm cr}=5/3$.
In the latter case (Fig.~\ref{fig:eps_vs_w}) $\epsilon > 0$ for $w_2>0$, up to maximum possible value
for $w_2$. However, for $\gamma_{\rm cr}=4/3$ $\epsilon$ first becomes negative for $w_2>0$, then reaches a minimum,
and then crosses again the line $\epsilon=0$. In other words for $\gamma_{\rm cr}=4/3$ there are for some Mach
numbers three solutions for $\epsilon=0$,
namely the standard shock solution (i.e. $w_2=0$), a solution that maximises $w_2$ and for which $\chi_{\rm sub}=1$, and a point somewhere in between these
two limits. These solutions correspond to the solutions of the two-fluid model of \citet{drury81}, which assumes
energy flux conservation.  $M_{\rm acc}$ corresponds to the Mach number where the two non-standard solutions coincide,
for which the sub-shock compression ratio is $\chi_{\rm sub}=5/2$ (Eq.~\ref{eq:chisubmax}).

For many astrophysical settings, especially in interplanetary shocks,  
for low Mach numbers the adiabatic index for the accelerated
particle population will more closely resemble $\gamma_{\rm cr}=5/3$. We illustrate this in Fig.~\ref{fig:kinetic}, which is not
based on the extended Rankine-Hugoniot relations of \citet{vink10a}, but on the semi-analytical kinetic solutions
of \citet{blasi05}. It shows that as the Mach number decreases $\gamma_{\rm cr}$ approaches 5/3. However,
the energy flux reaches $\epsilon=0$ for $M_{\rm g,0}\approx 2.79$, with a corresponding $\gamma_{\rm cr}\approx 1.57$, and $w_0\approx 0.15$.
For lower Mach numbers $\epsilon<0$.
Fig.~\ref{fig:mcrit} shows the critical Mach number for acceleration
as a function of the assumed adiabatic index for cosmic rays.

\subsection{Perpendicular, magnetically dominated shocks}
\label{sec:beta}
The best studied low Mach number shocks are arguably shocks
in the solar system. But these shocks often have a low upstream plasma-beta ($\beta_0 < 1$). 
The presence of significant pressure from a magnetic field component will make the flow less compressible,
and requires more work to be done by the shock in order to compress the plasma.
As a result, there will be less energy available for
accelerating particles.
Including
magnetic fields into the Rankine-Hugoniot solutions complicates the calculation
of shock parameters \citep{tidman71}, but one can obtain some insights
by considering the limiting case of a strictly perpendicular
shock in which all the upstream pressure is provided by the magnetic field;
so $\beta_0=0$,  $B_0=B_{0,\perp}$ and $P_{\rm g,0}=0$, and $w_0=0$.
The relevant shock equations are given in Appendix~\ref{sec:appendixb},
but here we list the main points.

For a strictly perpendicular shock with $\beta_0=0$, one finds for the shock compression ratio at the 
subshock (see Eq.~\ref{eq:chiperp})
\begin{equation}
\chi_{\rm sub}=-(M_{\rm A,1}^2+5/2)+\sqrt{D_{\rm 1}},\label{eq:chisubperp}
\end{equation}
with
\begin{equation}
D_{\rm 1}
\equiv M_{\rm A,1}^4+13M_{\rm A,1}^2+\frac{25}{4},
\end{equation}
with the numerical values valid for $\gamma_{\rm g}=5/3$.
The subshock Alfv\'en Mach number is given by
\begin{equation}
M_{\rm A,1}^2=M_{\rm A,0}^2\chi_{\rm prec}^{-3}.\label{eq:machsub}
\end{equation}

The maximum compression ratio can be found in analogy with the procedure
that lead to Eq.~\ref{eq:chiprecmax}, namely by determining $d\chi_{\rm tot}/d\chi_{\rm prec}=0$ in the limit
of $\chi_{\rm prec}=1$, with
\begin{align}
\chi_{\rm tot}=& \chi_{\rm prec}\chi_{\rm sub}\\ \nonumber
&
=-(M_{\rm A,0}^2\chi_{\rm prec}^{-2}+\frac{5}{2}\chi_{\rm prec})+\sqrt{
\chi_{\rm prec}^2D_{\rm 1}.
}
\end{align}
After some algebra one finds that in the limit $\chi_{\rm prec}\rightarrow 1$,
and  $M_{\rm A,0}=M_{\rm A,1}$, the solution has
to obey the relation
\begin{equation}
(8M_{\rm A,0}^2-10)\sqrt{D_{\rm 1}}-(8M_{\rm A,0}^4+26M_{\rm A,0}^2-25)=0.
\end{equation}
The solution to this equation is $M_{\rm A,0}=M_{\rm A, 1}=5/2$,
which corresponds to a subshock compression ratio of $\chi_{\rm sub}=5/2$
(Eq.~\ref{eq:chisubperp}).
So the critical Mach number for acceleration for a perpendicular shock with
$\beta_0=0$ and $w_0=0$ is $M_{\rm acc}=5/2$.

\begin{figure}
\centerline{
\includegraphics[angle=-90,width=0.5\textwidth]{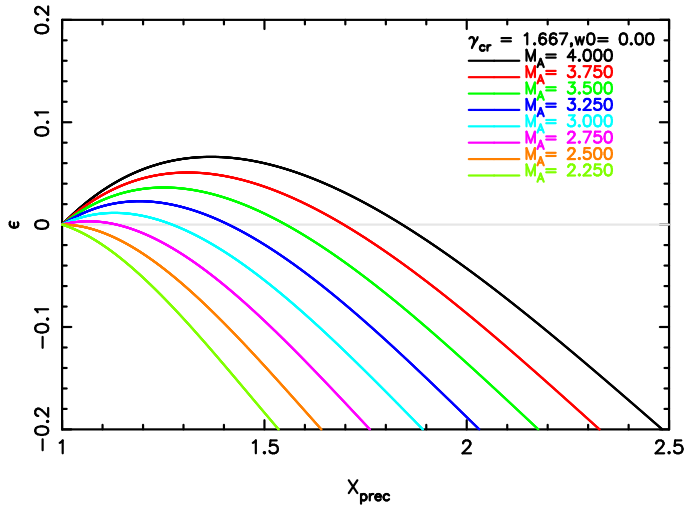}
}
\caption{
The same as Fig.~\ref{fig:betalarge} (left), but now for  perpendicular 
shocks with $\beta_0=0$, and with Mach numbers that include
the appropriate critical Alfv\'en Mach number $M_{\rm A}=2.5$ (orange).
\label{fig:perp}
}
\end{figure}

Eq.~\ref{eq:epsilonb} in Appendix \ref{sec:appendixb} is the equivalent of Eq.~\ref{eq:epsilon},
and shows which values of the compression ratio are allowed (i.e. $\epsilon \geq 0$).
The relation between $\epsilon$ and the precursor compression strength around $M_{\rm acc}$
is illustrated in Fig.~\ref{fig:perp}, 
which is similar to Fig.~\ref{fig:betalarge}.

\begin{figure}
\centerline{
\includegraphics[angle=-90,width=0.5\textwidth]{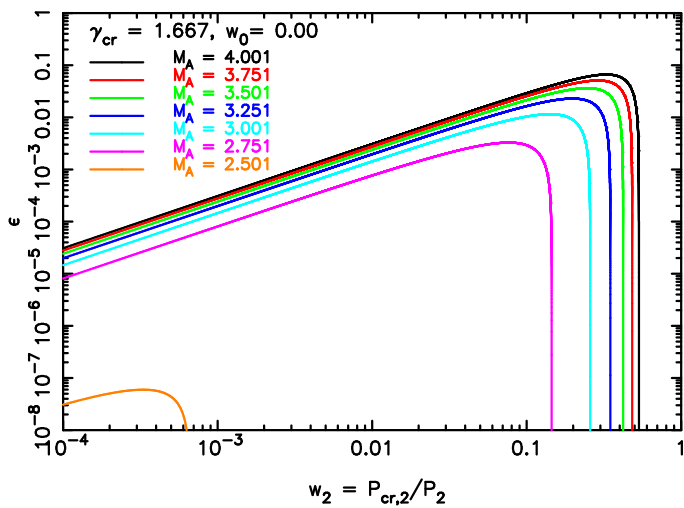}
}
\caption{
The same as Fig.~\ref{fig:eps_vs_w}, but now for  perpendicular 
shocks with $\beta_0=0$, and with Mach numbers that include
a value close to the critical Alfv\'en Mach number $M_{\rm A}=2.5$ (orange).
\label{fig:eps_vs_w_perp}
}
\end{figure}

In order to illustrate  the effects of the critical Mach number on particle acceleration, Fig.~\ref{fig:eps_vs_w_perp}
shows the possible three-fluid solutions for the shock conditions and acceleration efficiency, with the third
"fluid" being the magnetic field. These curves are calculated using the appropriate expression for
the efficiency parameter $w_2$, which is now defined as
\begin{equation}
w_2 \equiv  \frac{P_{\rm cr,2}}{P_{\rm g,2} + P_{\rm cr,2}+P_{\rm B,2}}.\label{eq:w2b}
\end{equation}
The expression for $w_2$ as function of the Mach number, and the total and subshock compression ratios
is
\begin{equation}
w_2=\frac{(1-\chi_{\rm prec}^2)+2M_{\rm A,0}^2\left(1-\frac{1}{\chi_{\rm prec}}\right)}
{1 + 2M_{\rm A,0}^2\left(1 - \frac{1}{\chi_{\rm tot}}\right)}.\label{eq:wperp}
\end{equation}
Note the similarity with Eq.~\ref{eq:w2}:
inserting $\gamma=2$ and $w_0=0$ in that equation and replacing
$M_{\rm g,0}$ with $M_{\rm A,0}$ gives the above expression.

The results in this section, therefore, show that 
due to a lower compressibility of  plasmas with  
dominant magnetic field pressures, more work
needs to be done to compress the plasma, and,
as a result, the critical (Alfv\'en) Mach number for forming a precursor
is higher than for $\beta_0>>1$, $M_{\rm acc}=5/2$.

It is assumed here that the magnetic field is passive. If, however, the magnetic field is amplified
due to cosmic-ray streaming, or some turbulent dynamo mechanism, the resulting value of $M_{\rm acc}$ will be higher,
in a similar way as non-adiabatic heating in the precursor results in larger values for $M_{\rm acc}$.

\begin{figure*}
\centerline{
\includegraphics[width=0.95\textwidth]{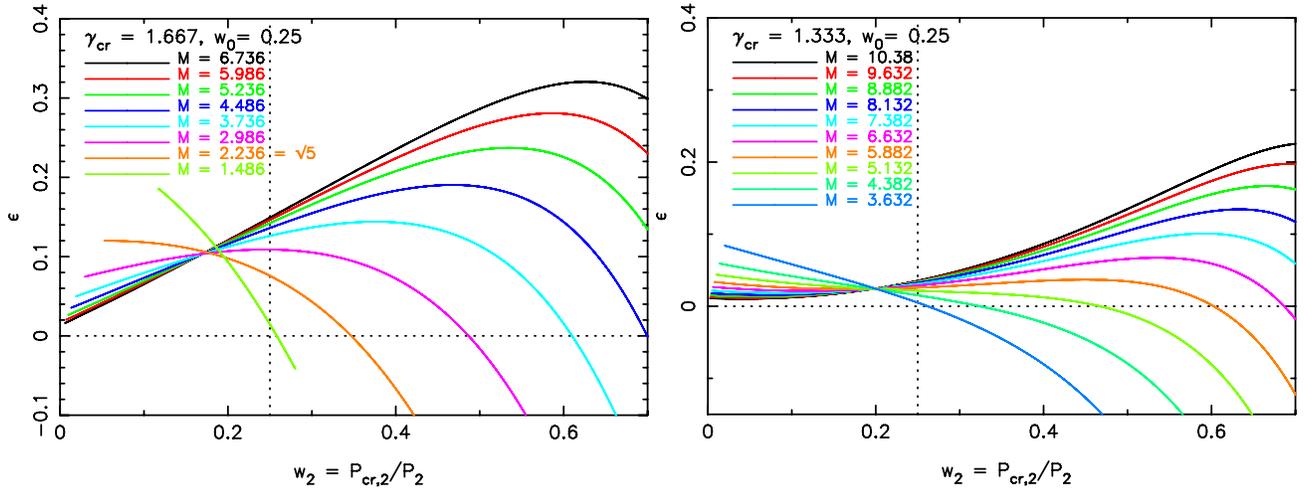}
}
\caption{
The solutions for the escape energy flux as a function of the downstream cosmic-ray pressure $w_2$ (similar to Fig.~\ref{fig:eps_vs_w}and \ref{fig:eps_vs_w_perp}), but now with the contribution of an additional upstream cosmic-ray pressure from pre-existing cosmic rays, $w_0=0.25$
(Eq.~\ref{eq:epsilon}).
The left panel is for a non-relativistic accelerated particle population ($\gamma_{\rm cr}=5/3$), the
right panel is for relativistically dominated particles ($\gamma_{\rm cr}=4/3$).
The Mach numbers differ 0.75 (1.33) times an integer number from the critical Mach number, $M_{\rm acc}=\sqrt{5}$ for $\gamma_{\rm cr}=5.3$
and $M_{\rm acc}=5.882$ for $\gamma_{\rm cr}=4/3$.
\label{fig:eps_vs_w_w0}
}
\end{figure*}

\begin{figure*}
\centerline{
\includegraphics[width=0.95\textwidth]{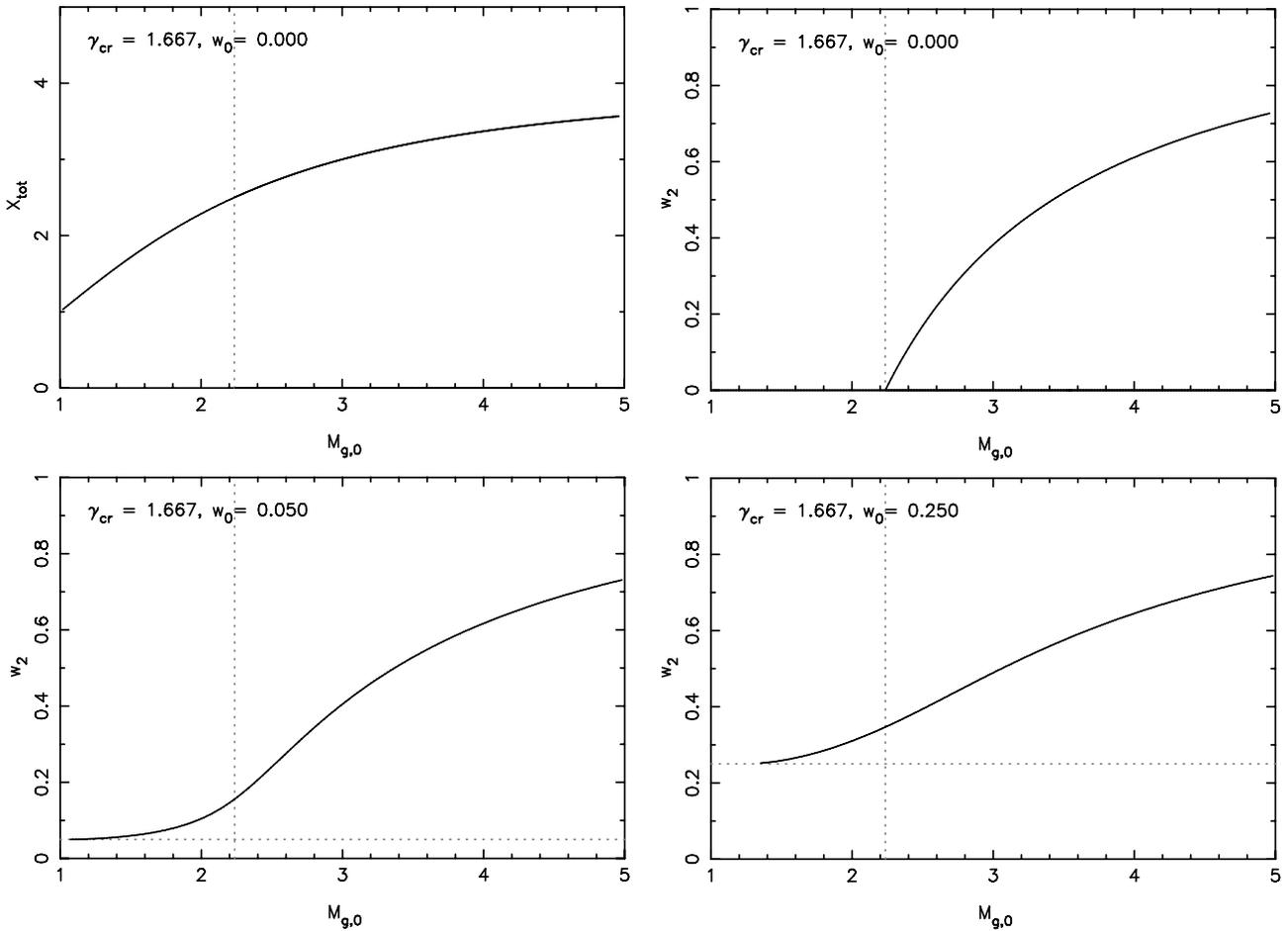}
}
\caption{
Shock solutions for $\gamma_{\rm cr}=5/3$ as a function of Mach number $M_{\rm g,0}$ for the case in which no energy
is escaping from the system ($\epsilon=0$), corresponding to the two-fluid model of \cite{drury81}.
Top left panel: the total compression ratio, which follows the standard Rankine-Hugoniot relations for $\gamma=5/3$.
Other panels: the downstream fractional cosmic-ray pressure for increasing values of the pre-existing cosmic-ray fractional pressure: 
$w_0=0, 0.05, 0.25$.
The vertical dotted line indicates the critical acceleration Mach number $M_{\rm acc}=\sqrt{5}$, whereas the horizontal dotted
line indicates $w_0$.
Note that the total compression ratios can be higher for $\epsilon > 0$, whereas the maximum values for $w_2$ provide upper bounds
for $\epsilon>0$.
\label{fig:w0_nonrel}
}
\end{figure*}

\begin{figure*}
\centerline{
\includegraphics[width=0.95\textwidth]{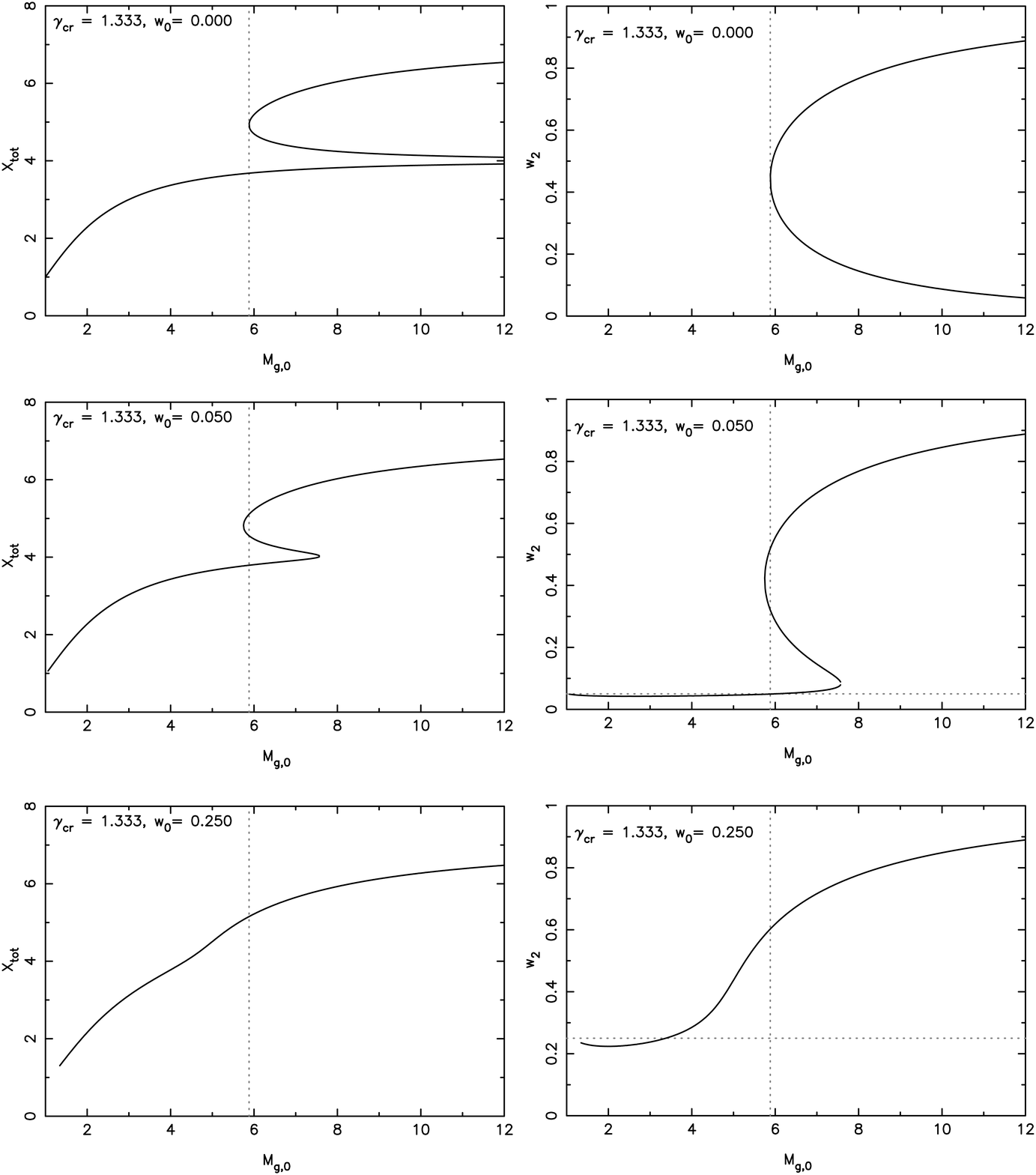}
}
\caption{
Similar to Fig.~\ref{fig:w0_nonrel}, but now for $\gamma_{\rm cr}=5/3$. The left hand panels show the total compression ratio $\chi_{\rm tot}$,
and the right hand panels the downstream fractional cosmic-ray pressure $w_2$, for increasing values of $w_0$. 
\label{fig:w0_rel}
}
\end{figure*}

\subsection{Shocks with pre-existing cosmic-rays}
\label{sec:w0}
In the solutions discussed above we assumed that there is no
population of pre-existing cosmic rays. However, pre-existing cosmic
rays can be incorporated in the extended Rankine-Hugoniot relations, by specifying
the additional parameter $w_0=P_{\rm cr,0}/P_0$, as explained in Appendix~\ref{sec:appendixa}.
The solutions to the energy flux equation (Eq.~\ref{eq:energy}) are shown in Fig.~\ref{fig:eps_vs_w_w0}
for non-relativistic ($\gamma_{\rm cr}=5/3$) and completely relativistic cosmic rays
($\gamma_{\rm cr}=4/3$).

These figures show that for $w_0>0$ it is possible to find solutions with $\epsilon \geq 0$ even for $M_{\rm g,0}<  M_{\rm acc}$.
However, some of these solutions are unphysical. For example, the left most limit of all the curves in the figures
correspond to no-precursor compression ($\chi_{\rm prec}=1$). 
The continuity of the cosmic-ray pressure in that case implies that
from far upstream to downstream the cosmic-ray pressure is constant
($P_{\rm cr,2}=P_{\rm cr,0}$). 
But it is impossible to have cosmic-rays take away
energy flux from the system, if there is no cosmic-ray pressure gradient 
present.\footnote{In fact, this could be a possible, but trivial
solution, if the pre-existing cosmic-rays do not couple to the gas at all.
In that case one should not write for the downstream enthalpy flux $H=[P_{\rm cr,2}+u_{\rm cr,2}+ P_{\rm g,2}+u_{\rm g,2}+\frac{1}{2}\rho_2v_2^2]v_2$,
but associate the cosmic rays still with the velocity of the upstream medium, as there is no coupling,
$H=[P_{\rm cr,2}+u_{\rm cr,2}]v_0+[P_{\rm g,2}+u_{\rm g,2}+\frac{1}{2}\rho_2v_2^2]v_2$.
In that case no escape flux is necessary for $w_0>0$ and $\chi_{\rm prec}>1$.
The problem arises that for $\chi_{\rm prec}=1,w_0>0$ the cosmic-ray pressure is continuous but leads nevertheless to an
 associated change in enthalpy flux, due to the change in frame velocity ($v_0 \rightarrow v_2$).
}

It is beyond the possibilities of the extended Rankine-Hugoniot relations to firmly state what parts of the curves with $w_0>0$ are physically possible.
Analytic solutions in the framework of the two-fluid model and $w_0>0$ do exist for the case of conservation of energy flux
\citep[$\epsilon=0$,][]{drury81,malkov96,becker01}, which correspond to the zero points in Fig.~\ref{fig:eps_vs_w_w0}.
These zero points are shown as a function of Mach number in Fig.~\ref{fig:w0_nonrel} and Fig.~\ref{fig:w0_rel}, for respectively
$\gamma_{\rm cr}=5/3$ and $\gamma_{\rm cr}=4/3$. They illustrate the different behavior for relativistic and non-relativistic accelerated
particles. 

For the non-relativistic case ($\gamma_{\rm cr}=5/3$), 
there is never more than one solution for $\epsilon=0$, if pre-existing cosmic rays are present ($w_0>0$).
For $w_0=0$ these solutions require $M_{\rm g,0}>M_{\rm acc}=\sqrt{5}$. The highest values for $w_2$  in case
we take energy flux conservation ($\epsilon=0$) provides an upper bound on
$w_2$ for solutions with escape (see Fig.~\ref{fig:eps_vs_w} and Fig.~\ref{fig:eps_vs_w_w0} (left)).
For completely relativistic cosmic rays ($\gamma_{\rm cr}=4/3$) there are for $w_0=0$ two solutions with $\epsilon=0$
and $w_2>0$. This leads to the bifurcation in $\chi_{\rm tot}$ and $w_2$ in the top  panels of Fig.~\ref{fig:w0_rel}
for $M_{\rm g,0}>M_{\rm acc}$.
Fig.~\ref{fig:w0_rel} once more illustrates that there is no solution with $w_0=0$ and $w_2>0$ for Mach numbers $M_{\rm g,0}<M_{\rm acc}\approx 5.88$.

Increasing the pressure in pre-existing cosmic rays  ($w_0>0$) changes the character of the solutions, as slowly the
bifurcation disappears, and also viable solutions exist for $M_{\rm g,0}<M_{\rm acc}\approx 5.88$.
The reason is that with a higher pressure in pre-existing cosmic rays, the shock solutions with $\epsilon=0$ start approaching the
standard Rankine-Hugoniot solutions for a relativistic gas, which for high Mach numbers approaches the compression ratio
$\chi_{\rm tot}=7$. Note that Fig.~\ref{fig:w0_rel} is similar  to the figures in \citet{malkov96}, showing that the extended
Rankine-Hugoniot relations explored  here encompass the  two-fluid model with conservation of energy flux \citep{drury81,malkov96,becker01}.

\section{Discussion}

\subsection{The case for a minimum Mach number for acceleration}
We showed that the ability to accelerate particles relies a critical magnetosonic Mach number $M_{\rm acc}$,
which depends on the presence/absence of perpendicular magnetic fields and the assumed
adiabatic index of the population of accelerated particles. If there are no pre-existing cosmic rays ($w_0=0$),
this critical Mach number is the minimum Mach number for which sufficient energy flux is available
to accelerate particles. In all cases the critical Mach number corresponds to a compression ratio at the sub-shock of
$\chi_{\rm sub}=5/2$, corresponding to a subshock Mach number $M_{\rm g,1}=\sqrt{5}$.
For non-relativistically dominated cosmic-rays the critical Mach number 
lies
in the range $\sqrt{5}\leq M_{\rm acc}\leq 5/2$, depending on whether
plasma beta is large, or very low. The values of $M_{\rm acc}$ are higher if
heating or magnetic field amplification are important, or if the non-thermal
particles have a significant relativistic component.  For completely relativistic cosmic
rays $M_{\rm acc}\approx 5.88$. 

The situation changes in case a pre-existing population of cosmic rays exist, in the sense
that in that case the additional degree of freedom allows for cosmic-ray acceleration
even for Mach numbers lower than $M_{\rm acc}$. However, not all the solutions found
with the extended Rankine-Hugoniot relations employed here,  may be physical possible,
because in some cases escape of energy flux is required, even though there are no
substantial pressure gradients in the cosmic rays.

The derivation of $M_{\rm acc}$ in the previous sections is based on only a few assumptions:
like for the general shock-jump relations, it relies on the plane parallel 
shock approximation; it requires steady state conditions; and it requires the subshock
to be governed by the standard Rankine-Hugoniot relations.

These assumptions are very generic and are common to most shock and diffusive shock 
acceleration models. However, the steady state assumption
leaves open the possibility that
particle acceleration is not a continuous phenomenon, 
but occurs irregularly or in bursts.

Another, more fundamental, issue is that if one observes the (sub)shock
region 
in detail the distinction between what is a precursor and what is the 
subshock becomes more complicated. 
We followed here the convention of
diffusive shock acceleration theories that refer to the main shock as the 
subshock.
However, in collisionless shock theory the subshock refers to the steep gradient
in density and pressure,  as opposed to other quantities, 
like magnetic field that may
change on slightly larger length scales.
Indeed, collisionless shocks, even with ignoring diffusive shock acceleration, can have 
a complex structure \citep{treumann09}. 
They have precompression in a so-called foot region, a steep shock ramp, 
a downstream overshoot region,
which corresponds to a compression ratio higher than allowed by the 
Rankine-Hugoniot relations, followed by an undershoot region. 
Only further downstream the flow relaxes to
the standard shock-jump conditions.
The foot region is associated with ions reflected immediately back 
upstream by the shock.
So the foot region could also be labeled a shock precursor. 
But, in the context of the discussion here, the
precursor/foot region should still be regarded as an integral part of the 
subshock itself.
The reason is that across the total subshock structure the standard
shock-jump relations are observed. The complex structure, and physical
processes like ion reflection, are a means by which nature forces the flow
to establish a shock and observe the Rankine-Hugoniot relations.
In contrast, shocks with diffusive shock acceleration do not observe
the Rankine-Hugoniot relations, and they can have compression
ratios much higher than the standard shock-jump relations. This is possible
due to the escape of high energy particles upstream.

Nevertheless, the distinction between an "accelerated particle precursor" and
a "foot region" may not be that sharp. The distinction is more easily defined
if shock acceleration is very efficient, and the accelerated particle
precursor becomes very extended. But around $M=M_{\rm acc}$ the efficiency
is low (Fig.~\ref{fig:eps_vs_w} and \ref{fig:eps_vs_w_perp}), and
it may observationally be difficult, or even arbitrary 
to distinguish between a precursor from diffusively accelerated particles 
and a foot region.

The appearance of foot regions, ion reflection, and overshoot regions is usually
associated with another critical Mach number,
the so-called {\em first critical Mach} number, $M_{\rm c}$, which has a range
of $1\leq M_{\rm c}\leq 2.76$, depending on the shock obliquity and plasma-beta \citep{edmiston84}, $M_{\rm c}=1$ corresponding to $\beta >> 1$ and 
$M_{\rm c}\approx 2.76$ corresponding to perpendicular shocks with $\beta_0=0$.

Below the first critical Mach number ordinary resistivity is sufficient to provide 
the necessary shock steepening, whereas for supercritical shocks anomalous dissipation
mechanisms are necessary to force the shock to observe the Rankine-Hugoniot relations.
Ion reflection is one of the ingredients by which the flow manages
to acquire the required shock heating. 
Indeed, ion reflection is observationally associated
with supercritical shocks, although some subcritical shocks also appear to have ion reflection
and overshoot regions \citep{mellot87}.
Note that the presence of an overshoot seems to violate the flux conservation laws (Eq.~\ref{eq:chi}-\ref{eq:energy}),
but this may be an indication that energy flux is temporarily stored in the electrostatic oscillations,
and therefore the equation of state is temporarily altered, corresponding to a lower specific heat ratio $\gamma$,
and higher compression ratios \citep{eselevich84}.

The idea that two critical Mach numbers may operate in the same Mach number regime
is interesting and may have some observational consequences. For
high beta shocks, the first critical Mach number is very low, $M_{\rm c}\approx 1$, and
lies below the critical Mach number for acceleration $M_{\rm acc}=\sqrt{5}$, hence $M_{\rm c}<M_{\rm acc}$. 
In contrast, for
very low beta, perpendicular shocks the first critical Mach number is $M_{\rm c}\approx 2.76$, which is
larger than $M_{\rm acc}\approx 2.5$.
The effects of the two different critical Mach numbers, $M_{\rm c}$ and $M_{\rm acc}$, may therefore be observationally investigated 
by exploiting this difference between low and high beta shocks.

\subsection{Comparison to observations}
Observationally the case for whether there is a critical Mach number for
particle acceleration is not so clear.
The Earth' bowshock is generally associated with Mach numbers above the
critical regime \citep[$M_{\rm ms}\approx 5$,][]{bale03}.
The solar wind termination shock has a Mach number in the range
where one may expect to see critical behavior \citep[$M_{\rm ms}\approx 2.5$,][]{lee09}.
\citet{florinski09} made a case for non-linear particle 
acceleration at the solar wind
termination shock, as  Voyager 2 data indicate the presence of a precursor
induced by accelerated particles.
The total compression ratio for that case was $\chi=3.1$, which is above the critical
value of $\chi_{\rm tot}=5/2$.

CMEs are also associated with particle acceleration, and Type II radio bursts
are  considered to be evidence for acceleration.
\citet{gopalswamy10} showed that Type II radio bursts are associated with high 
velocity/high Mach number CMEs (with mean velocities of 1237 km\,s$^{-1}$ ) and the radio quiet CMEs with low
velocities (with mean velocities of 537  km\,s$^{-1}$).
The Mach numbers of the low velocity CMEs were still relatively high, with
a median of $M_{\rm ms}=2.3$ and an average of $M_{\rm ms}=2.7$. The latter value is above
the critical Mach number derived here, and close to the first critical Mach number $M_{\rm c}$.
But it should be noted that the errors on the Mach numbers are relatively high \citep[systematic error $\Delta M\approx 0.55$,][]{gopalswamy10}.
\citet{pulupa10} even concluded that the measured Mach numbers are not well correlated
with the occurrence of Type II radio bursts, whereas there is a strong correlation with
velocity. 

Another measure for the compression ratio for shocks associated with Type II radio bursts is
the bandwidth of the radio emission. The work by \citet{mann95} indicates that
the minimum bandwidth is $\Delta f/f=0.16$, which, according to \citet{mann95b},
corresponds to a minimum shock-compression ratio of $\chi=1.35$. This is clearly not in accordance
with the critical Mach number $M_{\rm acc}$ derived in the present paper, which occurs
for a compression ratio of 2.5 or more. However,
it is not clear yet whether the bandwidth is indeed caused by the jump in the density caused by the shock,
or whether density gradients in the upstream region are responsible.
A joint analysis of the location of the radio emission and optical CME locations seems
to suggest that the radio emission is in general coming from a region upstream of the shock \citep{ramesh12}.

Clearly, the uncertainty of the correlation between Type II bursts and Mach numbers
could be resolved by more precise measurements
of the Mach numbers, rather than the shock velocity, for those exact locations
that emit in the radio. A recent analysis of SOHO observations by 
\citet{bemporad11} shows that more precise Mach numbers can be obtained, indicating
that the highest compression ratios, $\chi\approx 3$, are found near the center of the CME.
A problem may remain that for CMEs the plasma beta is rather low, so that the determination
which critical Mach number determines Type II bursts, $M_{\rm c}$ or $M_{\rm acc}$, may
be difficult to distinguish.

For this reason it is very interesting that recently \citet{giacalone12} showed that all shocks that have high enough compression
ratios show evidence for particle acceleration. Interestingly, this study uses as an indication
of a strong shock a compression ratio of $\chi\geq 2.5$, which is exactly the compression
ratio associated with lowest possible value for the critical Mach number $M_{\rm acc}=\sqrt{5}$
in case of a sonic shock, and $M_{\rm acc}=2.5$ for a magnetically dominated, perpendicular shock.

Apart from Mach number, another factor that appears to influence the
presence or absence of accelerated particles associated with CMEs is the
occurrence of a CME preceding the event by less than a day 
\citep{kahler99,gopalswamy04}. This correlation has been attributed to
the presence of non-thermal particle populations created by the first CME
\citep{laming13}. Our theoretical results here indicate that the mere
presence of accelerated particles may facilitate particle acceleration
for Mach numbers lower than the critical Mach number. Note that both effects,
the influence on the jump relations, and the presence of seed-particles,
may play complementary roles.

In this context one should raise the question to what extent the
omnipresent Galactic cosmic rays are important. This likely depends on
the length scale of the coupling between cosmic rays
and the plasma directly up- and downstream of the shock. If the length
scale is much longer than the typical length scales over which the shock
develops, these pre-existing cosmic rays are likely to not
affect the shock structure.
For that reason, for CMEs probably only low 
energy accelerated particles are important
(keV to MeV energies). So particles from preceding CMEs are much more
important than Galactic cosmic rays. However, these are subtleties that require
further investigation.

The largest shocks observed in the Universe are those in clusters of galaxies.
Many of them are detected as discontinuities in the X-ray emission \citep{markevitch07}. 
These shocks
are caused by infalling subclusters or galaxy groups, or due to mergers of clusters.
Some shocks are detected through their non-thermal radio emission, clearly
indicating that at these shocks electrons are accelerated \citep{vanweeren10,hoeft11}.
The radio detected shocks, often called radio relics, are usually located in the outskirts of the cluster.
The shock velocities can be several thousand km\,s$^{-1}$, but due to the high plasma
temperatures, $kT\approx 1-10$~keV, the Mach numbers are usually modest $M_{\rm ms}\lesssim 3$.
The radio relics are mostly found in the periphery of the clusters where the density is lower
than in the center, whereas the magnetic field may be as  high as a few $\mu$G.
The plasma betas are believed to be $\beta \approx 1-10$ (Markus Hoeft, private communication).
The lack of radio emission from many X-ray detected shocks suggest that there
is, indeed, a dependence of radio emission on Mach number, which 
could therefore hint at the existence of a critical Mach number for acceleration.
It is usually assumed that
the onset of radio emission happens in the range of $2<M_{\rm acc}<3$ \citep{hoeft11}.\footnote{
These exact Mach numbers are not easily measured, and either rely on interpreting the
radio spectrum in the context of test particle acceleration, or on the detection of the
shock in X-rays. However, it is not always clear whether the X-ray detected shock and
the shock associated with the radio emission exactly coincide \citep{ogrean13}.}
This should be contrasted to the first critical Mach number, $M_{\rm c}$, which in clusters of galaxies is likely smaller than 2.
Therefore, the critical Mach number derived in the present paper may be important
for the presence or absence of radio emission from shocks in clusters of galaxies. 
However, the derived numbers for $M_{\rm acc}$ 
were for non-relativistic particles. The radio emission
is caused by relativistic electrons. 
As long as the protons are non-relativistic and dominate the population of
accelerated particles, $\gamma_{\rm cr}=5/3$, 
may still be a reasonable approximation. 
If protons are accelerated
to relativistic energies, with $E>938$~MeV, $\gamma_{\rm cr}$ will decrease 
toward  $\gamma_{\rm cr}=4/3$, and $M_{\rm acc}$ will increase. 
As discussed in Sect.~\ref{sec:highbeta}, it depends on the
spectral energy distribution what the effective specific heat ratio of 
the accelerated particles is. 
But for a significant component of relavistic protons a
limiting Mach number of $M_{\rm acc}\approx 3$ is likely. 
This could mean that many of the observed relics cannot accelerate protons
to very high energies, and only the highest Mach number shocks ($M>3$)
contain significant fractions of relativistic protons. 

Alternatively, the limiting Mach numbers for shocks moving through a medium
containing cosmic rays is more relaxed (Eq.~\ref{sec:w0}). So evidence for
relativistic particles associated with low Mach number shocks, may indicate
the presence of pre-existing cosmic rays in the intra-cluster medium.
As is the case for CME induced shocks, for clusters the importance
for pre-existing cosmic rays as seed particles for further acceleration
has been pointed out. And also in this case it should
be pointed out that pre-existing cosmic rays may have two, complementary, 
effects:
it changes the degrees of freedom of the shock system, allowing for
acceleration for lower Mach numbers (the present work), and it may
help as a source of seed particles, which are injected in the shocked
and then experience further acceleration \citep{pinzke13}.

Another effect could be that
acceleration becomes discontinuous: for $\sqrt{5}<M< 3$ particles are being 
accelerated but once a significant number of protons become relativistic
the acceleration efficiency goes dramatically down for some time,
and then start up again.
Clearly these effects need to be further investigated, both observationally
in shocks close the critical Mach number, 
and with more elaborate  kinetic shock-acceleration models.

\section{Conclusion}

We presented in this paper a derivation of a critical Mach number for particle
acceleration, $M_{\rm acc}$. The basic idea is that diffusive shock acceleration is
inherently non-linear, and results in the compression and slowing down of the upstream plasma,
forming a so-called shock precursor.
It turns out that adiabatic compression in the precursor followed by
a shock, as given by the standard shock jump conditions, cannot be energetically
sustained for Mach numbers smaller than a critical value $M_{\rm acc}=\sqrt{5}$.
This limit is even higher for magnetic dominated plasmas, which in the extreme
case of $\beta_0=0$ and purely perpendicular shock gives a critical Mach number of
$M_{\rm acc}=2.5$.
In case there is substantial
 pre-existing cosmic-ray population the limits on further acceleration may be relaxed.
This critical Mach number  should not be confused with the so-called first
critical Mach number, which, depending on obliquity and $\beta_0$, lies in the
range $1 \leq M_{\rm c} < 2.76$ \citep{edmiston84}. 

We discussed the critical Mach number, $M_{\rm acc}$,
in connection with observational
evidence for particle acceleration at low Mach number shocks, such as
in the solar system or in clusters of galaxies, and in conjunction
with first critical Mach number. There is indeed observational evidence for
a Mach number dependence of particle acceleration with Mach number, which
agrees with the idea that between Mach numbers of 2-3 the acceleration
properties of shocks change. However, the observational evidence is not
precise enough to judge whether there is indeed a critical Mach number range
for acceleration
$\sqrt{5}<M_{\rm acc}<2.5$, or whether the observed phenomenology of solar system
shocks is governed by the first critical Mach number $M_{\rm c}$.

For shocks in clusters of galaxies, there is some indication that Mach numbers
above $2-3$ are needed to create a population of radio synchrotron emitting electrons.
It is pointed out that the critical Mach number, $M_{\rm acc}$, increases if 
the energetics
of the accelerated particles are dominated by relativistic particles,
which could mean that there is a strong
limit on the number fraction of relativistic protons in cluster shocks
with Mach number $M <3$.

\acknowledgements
It is a pleasure to thank Stefano Gabici for useful discussions.
The writing of this paper was stimulated by discussions during the 
JSI Workshop "Nature's Particle Accelerators", held in October 2012.
I thank the organizers for inviting me to this stimulating workshop.
I also thank Matthias Hoeft for discussions on shocks in clusters of galaxies.

\appendix

\section{The extended Rankine-Hugoniot relations including pre-existing cosmic rays}
\label{sec:appendixa}
                                   
\citet{vink10a} described a version of the Rankine-Hugoniot relations extended
with a component of accelerated particles. 
Like the Rankine-Hugoniot relations it evaluates the mass, momentum, and enthalpy
flux, but with some modifications:
Instead of applying the relations to two regions (upstream
and downstream of the shock) the relations are evaluated 
at three specific locations:
0) the (undisturbed) far upstream medium, 1) in the cosmic-ray shock precursor,
just upstream of the subshock (i.e. the actual gas shock), 
and (2) downstream of the subshock.  The standard Rankine-Hugoniot relations 
only consider (0) and (2).
Unlike the standard Rankine-Hugoniot relations we allow energy flux to escape
from the overall system, which is a standard outcome of 
kinetic models of cosmic-ray acceleration \citep[][for an overview]{caprioli10}.
The system can be closed
using the condition that the gas pressure does have a shock-jump at the 
sub-shock, but the cosmic-ray pressure ($P_{\rm c}$) is continuous across the shock,
which is a necessary consequence of diffusive shock acceleration \citep[see for example the appendix of][]{becker01}, i.e $P_{\rm cr,1}=P_{\rm cr,2}$.
It is important to note that in the context of this model the continuity of cosmic-ray pressure across the subshock is 
what sets the cosmic-ray component apart from the gas component.

For a given upstream gas Mach number $M_{\rm g,0}$, and an assumed adiabatic index, $\gamma_{\rm cr}$, for the
cosmic-ray component, the extended Rankine-Hugoniot relations give a range
of solutions that can be parametrized by the cosmic-ray precursor compression ratio $\chi_{\rm prec}\equiv\rho_1/\rho_0$.
The standard Rankine-Hugoniot shock jump solutions are retrieved for $\chi_{\rm prec}=1$.

Here we summarize the solutions presented in \citet{vink10a}, but augmented  with an additional parameter,
namely the upstream cosmic-ray pressure ($P_{\rm cr, 0}$). We do this by extending the use of the fractional
cosmic-ray  pressure \footnote{This is denoted $N$ in \citet{drury81}. Note that
\citet{becker01} uses the upstream cosmic-ray Mach number,
defined as $M_{\rm cr,0}=\sqrt{\rho_0V_s^2/\gamma_{\rm cr}P_{\rm cr,0}}$.
The relation between $w_0$ and $M_{\rm cr,0}$ is 
$w_0=1/(1+\gamma_{\rm cr}M_{\rm cr,0}^2/\gamma_{\rm g}M_{\rm g,0}^2)$.
},
\begin{equation}
w\equiv\frac{P_{\rm cr}}{P_{\rm g}+P_{\rm cr}},\label{eq:w}
\end{equation}
to the upstream region. The subscript "g" refers to the gas (thermal) component.
So $w$ in \citet{vink10a} is now labeled $w_2$ and the upstream quantity is $w_0$.

The conservation of mass flux  ($\rho v$) and momentum flux
($P_{\rm cr}+P_{\rm g}+\rho v^2$) throughout the whole shock
system can be made dimensionless by dividing pressure by the upstream
ram pressure $\rho_0 V_{\rm s}^2$, with $V_{\rm s}(=v_0)$ the shock velocity,
and using the compression factors
\begin{align}
\chi_{\rm prec}=\frac{\rho_1}{\rho_0}=\frac{v_0}{v_1}, \  \chi_{\rm sub}=&\frac{\rho_2}{\rho_1}=\frac{v_1}{v_2}, \ 
\chi_{\rm tot}=\chi_{\rm prec}\chi_{\rm sub}=\frac{\rho_2}{\rho_0}=\frac{v_0}{v_2},\label{eq:chi}
\end{align}
which express mass flux conservation.

To make momentum flux conservation dimensionless it is convenient
to use the definition of the gas Mach number
\begin{align}
M_{\rm g,0}\equiv &\sqrt{\frac{\rho_0V_{\rm s}^2}{\gamma_{\rm g}P_{\rm g,0}}}=\frac{V_{\rm s}}{c_{\rm sound}},\\
M_{\rm g,1}\equiv &\sqrt{\frac{\rho_1v_{1}^2}{\gamma_{\rm g}P_{\rm g,1}}}=\frac{v_{1}}{c_{\rm sound}}=M_{\rm g,0}\chi_{\rm prec}^{-(\gamma_{\rm g}+1)/2},\label{eq:m1}
\end{align}
with Eq.~\ref{eq:m1} indicating that we assume that the compression
of the gas in the precursor (region 1) is purely adiabatic.

The dimensionless pressures $\mathcal{P}_i$ ($i=0,1,2$) are then given by the following
relations
\begin{align}
&\mathcal{P}_0\ \equiv \frac{P_{\rm g,0}+P_{\rm cr,0}}{\rho_0 V_s^2}=
\frac{1}{\gamma_{\rm g}M_{\rm g,0}^2}\Bigl(\frac{1}{1-w_0}\Bigr),\\
&\mathcal{P}_1\ \equiv \frac{P_{\rm g,1}+P_{\rm cr,1}}{\rho_0 V_s^2}=  \frac{1}{\gamma_{\rm g}M_{\rm g,0}^2}\Bigl(\frac{1}{1-w_0}\Bigr)+ \Bigl(1-\frac{1}{\chi_{\rm prec}}\Bigr),\\
&\mathcal{P}_{\rm g,1}\equiv \frac{P_{\rm g,1}}{\rho_0V_{\rm s}^2}=\frac{\chi_{\rm prec}^{\gamma_{\rm g}}}{\gamma_{\rm g}M_{\rm g,0}^2},\label{eq:p1gas}\\
&\mathcal{P}_2\ \equiv \frac{P_2}{\rho_0V_{\rm s}^2}=\frac{1}{\gamma_{\rm g}M_{\rm g,0}^2}\Bigl(\frac{1}{1-w_0}\Bigr)+ \Bigl(1 -\frac{1}{\chi_{\rm tot}}\Bigr),\label{eq:p2}\\
&\mathcal{P}_{\rm g,2}\equiv \frac{P_{\rm g,2}}{\rho_0 V_s^2}=(1-w_2)\mathcal{P}_2=\frac{\chi_{\rm prec}^{\gamma_{\rm g}}}{\gamma_{\rm g}M_{\rm g,0}^2}+\Bigl(1 - \frac{1}{\chi_{\rm sub}}\Bigr)\frac{1}{\chi_{\rm prec},},\label{eq:p2gas}\\
&\mathcal{P}_{\rm cr,2}= \mathcal{P}_{\rm cr,1}=w_2\mathcal{P}_2.
\end{align}
Eq.~\ref{eq:p2gas} follows from the relation $P_{2}=P_{1}+(1-1/\chi_{\rm sub})\rho_1 v_1^2$, which
is similar to Eq.~\ref{eq:p2}.

The fractional pressure of cosmic-rays downstream $w_2$ can be derived from combining Eq.~\ref{eq:p2} and Eq.~\ref{eq:p2gas},
\begin{equation}
w_2= \frac{1 - (1-w_0)\chi_{\rm prec}^{\gamma_{\rm g}} +
(1-w_0)\gamma_{\rm g} M_{\rm g,0}^2\Bigl(1 - \frac{1}{\chi_{\rm prec}}\Bigr)}
{1 + 
(1-w_0)\gamma_{\rm g} M_{\rm g,0}^2 \Bigl(1 -\frac{1}{\chi_{\rm tot}}\Bigr)
}.\label{eq:w2}
\end{equation}
Setting $w_0=0$ (i.e no upstream cosmic rays) gives the expression found by \citet{vink10a}, and its asymptotic approximation
($M_{\rm g,0}\rightarrow \infty,  w_0=0$) is $w_2\approx (\chi_{\rm tot}-\chi_{\rm sub})/(\chi_{\rm tot}-1)$.

To complete the set of equations we give here the sub-shock compression ratio, which is simply the standard
Rankine-Hugoniot relation, applied to the gas component in region 1  \citep{malkov01,becker01,blasi05}:
\begin{equation}
\chi_{\rm sub}=\frac{(\gamma_{\rm g} + 1)M_{\rm g,1}^2}{(\gamma_{\rm g}-1)M_{\rm g,1}^2 + 2}.\label{eq:chisub}
\end{equation}

Equation \ref{eq:chi} to \ref{eq:chisub} are sufficient to predict
all shock relations, and cosmic-ray contributions, for a given value
of the main variable,  $\chi_{\rm prec}$, the precursor compression ratio.
In case that $w_0=0$, or $w_2 >>w_0$, $w_2$ provides 
a direct measure for the
cosmic-ray acceleration efficiency. But in order to see whether the solutions are physically possible we need to evaluate 
whether the enthalpy flux ($[P+u+\frac{1}{2}\rho v^2]v$) is either conserved, or energy is leaking out of the system
by escaping cosmic rays. In dimensionless form (i.e. dividing enthalpy by $\frac{1}{2}\rho_0 V_s^3$) we can express enthalpy
(non-)conservation as
\begin{align}
\left\{
\frac{\gamma_{\rm g}}{\gamma_{\rm g}-1}\mathcal{P}_{\rm g,2}  +
\frac{\gamma_{\rm cr}}{\gamma_{\rm cr}-1}\mathcal{P}_{\rm cr,2}  
+ \frac{1}{2}\frac{1}{\chi_{\rm tot}}\right\}\frac{1}{\chi_{\rm tot}} = 
\left\{
\frac{\gamma_{\rm g}}{\gamma_{\rm g}-1}\mathcal{P}_{\rm g,0}  +
\frac{\gamma_{\rm cr}}{\gamma_{\rm cr}-1}\mathcal{P}_{\rm cr,0}  +
(1-\epsilon)
\frac{1}{2}\right\},\label{eq:energy}
\end{align}
with $\epsilon\geq 0$, with $\epsilon=0$ indicating enthalpy conservation
\citep[c.f.][]{berezhko99,malkov01}.\footnote{
We take here that the escaping energy flux cannot exceed the free energy flux of the system ($\frac{1}{2}\rho V_s^3$).
}

If we write for convenience \footnote{In principle the adiabatic index of the cosmic rays upstream may
differ from that downstream, but we assume the cosmic rays are characterized by a unique number,
$4/3\leq \gamma_{\rm cr} \leq 5/3$.}
\begin{equation}
G_0\equiv w_0 \frac{\gamma_{\rm cr}}{\gamma_{\rm cr}-1}+(1-w_0) \frac{\gamma_{\rm g}}{\gamma_{\rm g}-1},
G_2\equiv w_2 \frac{\gamma_{\rm cr}}{\gamma_{\rm cr}-1}+(1-w_2) \frac{\gamma_{\rm g}}{\gamma_{\rm g}-1},
\end{equation}
Eq.~\ref{eq:energy} can with the help of Eq.~\ref{eq:p2} be  rewritten as
\begin{equation}
\epsilon = 1 +
\frac{2}{\gamma_{\rm g}M_{\rm g,0}^2}\Bigg(\frac{1}{1-w_0}  
\Bigg)
\Bigg[G_0 -\frac{G_2}{\chi_{\rm tot}}
\Bigg]
-\frac{2G_2}{\chi_{\rm tot}} + 
\frac{1}{\chi_{\rm tot}^2}(2G_2-1).
\label{eq:epsilon}
\end{equation}

\section{Shock solutions for perpendicular shocks}
\label{sec:appendixb}
In the limit of an upstream plasma that is dominated by magnetic pressure, i.e.
$\beta_0 \approx 0$ and $w_0=0$, one can ignore the upstream gas pressure $P_{\rm g,0}$ and precursor gas pressure $P_{\rm g,1}$ in
Eq.~\ref{eq:p2} and \ref{eq:energy}, but instead one has to
         introduce the pressure  caused by the perpendicular
magnetic field component. 
Hence, the momentum flux
conservation equation for a perpendicular, magnetically dominated, shock  is approximated by
\begin{equation}
\frac{B_{\perp,0}^2}{8\pi}+\rho_0V_{\rm s}^2=
P_1 + \frac{B_{\perp,1}^2}{8\pi}+\rho_1v_1^2 =
P_2 + \frac{B_{\perp,2}^2}{8\pi}+\rho_2v_2^2,
\label{eq:momentumperp}
\end{equation}
with $P=P_{\rm g}+P_{\rm cr}$ referring to particle induced pressure only (thermal and non-thermal).

These equations can be normalized using the Alfv\'en Mach number 
$M_{\rm A,0}\equiv V_{\rm s}/V_{\rm A}=V_{\rm s}/(B_{\perp,0}/\sqrt{4\pi\rho_0})$,
using the relation
\begin{equation}
\mathcal{P}_0= 
\frac{P_0}{\rho_0 V_{\rm s}^2}=
\frac{1}{2M_{\rm A,0}^2}.
\end{equation}
Here and in what follows $\mathcal{P}$ refers to the total pressure,
including the contribution of the magnetic field.
Using the above relations, we find that
\begin{equation}
\mathcal{P}_2=
\frac{1}{2M_{\rm A,0}^2}+
\Bigl(1-\frac{1}{\chi_{\rm tot}}\Bigr).\label{eq:momentumperpB}
\end{equation}
The pressure of the accelerated particles is on both sides of the subshock
assumed to be equal, hence $P_{\rm cr,2}=P_{\rm cr,1}=w_2(P_2 + B_{\perp,2}^2/(8\pi))$, 
with $w_2$ defined in Eq.~\ref{eq:w2b}.
Together with Eq.~\ref{eq:momentumperp} this means that
\begin{equation}
\mathcal{P}_{\rm cr,1}=\mathcal{P}_{\rm cr,2}=
w_2\Bigl[
\frac{1}{2M_{\rm A,0}^2}+\Bigl(1-\frac{1}{\chi_{\rm tot}}\Bigr)
\Bigr].\label{eq:pcr}
\end{equation}

Assuming only adiabatic compression of the magnetic
field, with $B_{\perp,1}=\chi_{\rm prec}B_{\perp,0}$ and
$B_{\perp,2}=\chi_{\rm tot}B_{\perp,0}$
\footnote{
Note that magnetic field amplification may be important
for strong Mach number shocks \citep[see][for observational and theoretical reviews]{helder12,schure12}.}, and using the fact that $P_{\rm cr,1}=P_{\rm cr,2}$
one can relate the downstream thermal pressure to the
pressure in the precursor, which gives
\begin{equation}
\mathcal{P}_{\rm g,2}
=\frac{\chi_{\rm prec}^2-\chi_{\rm tot}^2}{2M_{\rm A,0}^2}
+\frac{1}{\chi_{\rm prec}}\Bigl(1-\frac{1}{\chi_{\rm sub}}\Bigr).
\label{eq:p2therm}
\end{equation}
Comparing this with Eq.~\ref{eq:momentumperpB} shows that this should
be equal to 
\begin{equation}
\mathcal{P}_{\rm g,2}=
\mathcal{P}_2-\frac{\chi_{\rm tot}^2}{2M_{\rm A,0}^2}-\mathcal{P}_{\rm cr,2}
=
- \frac{\chi_{\rm tot}^2}{2M_{\rm A,0}^2}+
(1-w_2)\Bigl[
\frac{1}{2M_A^2}+\Bigl(1-\frac{1}{\chi_{\rm tot}}\Bigr)\Bigr],\label{eq:pth}
\end{equation}
which states that the downstream thermal pressure is the total pressure
minus the partial pressures of the magnetic field and the accelerated particles
(Eq.~\ref{eq:pcr}).
Combining Eq.~\ref{eq:pth} and ~\ref{eq:p2therm} one arrives at Eq.~\ref{eq:wperp}, given in the main text.

Finally, in order to complete the set of equation one needs to know the 
compression factor of a perpendicular, $\beta_0=0$, shock as a function of
Alfv\'en Mach number. 

In order to determine the shock compression ratio for a perpendicular
shock with $\beta_0=0$ one has to solve the enthalpy flux equation,
\begin{equation}
\frac{1}{2}\rho_2v_2^3+G_2P_2v_2+v_2\frac{B_{\perp,2}^2}{4\pi}=
(1-\epsilon) \frac{1}{2}\rho_0V_S^3 + \frac{V_SB_{\perp,0}^2}{4\pi}.
\label{eq:enthalpyB}
\end{equation}
Substituting Eq.~\ref{eq:momentumperp} into Eq.~\ref{eq:enthalpyB}, one
can find the following expression for energy escape
\begin{align}
\epsilon=& 1+\frac{2}{M_{\rm A,0}^2} -\frac{2\chi_{\rm tot}}{M_{\rm A,0}^2}
-\frac{G_2}{\chi_{\rm tot}M_{\rm A,0}^2}(1-\chi_{\rm tot}^2)
-\frac{2G_2}{\chi_{\rm tot}}\Bigl(1-\frac{1}{\chi_{\rm tot}}\Bigr)
           -\frac{1}{\chi_{\rm tot}^2},\label{eq:epsilonb}
\end{align}
with $G_2$ as defined under Eq.~\ref{eq:epsilon}. 
This equation is the equivalent for Eq.~\ref{eq:epsilon}, 
but now for perpendicular shocks,with $\beta=0$. 

The standard Rankine-Hugoniot solution, corresponding to $\epsilon=0$, can 
be found by solving the following cubic equation
\begin{equation}
(G-2)\chi^3
+(M_{\rm A}^2+2)\chi^2
-G(2M_{\rm A}^2+1)\chi
+(2G-1)M_{\rm A}^2 =0,\label{eq:solve}
\end{equation}
where the subscripts have been dropped, as this is a general shock-jump condition for
a perpendicular shock with $\beta_0=0$.
Eq.~\ref{eq:solve} has one trivial solution, $\chi=1$, which helps to transform 
the
cubic equation into a quadratic equation, which has one non-negative solution
\begin{equation}
\chi=\frac{-(M_{\rm A}^2+G)+\sqrt{D}}{2(G-2)} = 
 -\left(M_{\rm A}^2+\frac{5}{2}\right) + \sqrt{D},\label{eq:chiperp}
\end{equation}
with
\begin{equation}
D\equiv M_{\rm A}^4-18G_2M_{\rm A}^2+8G^2M_{\rm A}^2+8M_{\rm A}^2+G^2=
M_{\rm A}^4+13M_{\rm A}^2+\frac{25}{4},
\end{equation}
with the numerical values found by using $\gamma=5/3$, which gives $G=5/2$.
Asymptotically $\chi\rightarrow 4$ for $M_{\rm A} \rightarrow \infty$, which is the
shock jump condition for a strong shock.

This solution can also be used for the subshock, using $G=\gamma_{\rm g}/(\gamma_{\rm g}-1)=5/2$ and the Alfv\'enic Mach number
at the sub-shock (c.f. Eq.~\ref{eq:machsub}),
\begin{equation}
M_{\rm A,1}^2=\frac{1}{2}\frac{\rho_1v_1^2}{B_{\perp,1}^2/8\pi}=M_{\rm A,0}^2\chi_{\rm prec}^{-3}.
\end{equation}

\end{document}